\documentclass[12pt,preprint]{aastex}
\slugcomment{To appear in Ap.J. (Letters)}

\shorttitle{Black hole mass of BL Lacs}
\shortauthors{Falomo et al.}

\def\ref{\par \noindent \hang}

\begin{document}

\title{The black hole mass of BL Lac objects from the stellar velocity 
dispersion of the host galaxy}

\author{R. Falomo}
\affil{Osservatorio Astronomico di Padova, Vicolo dell'Osservatorio 5, 
35122 Padova, Italy}
\email{falomo@pd.astro.it}

\author{J.K. Kotilainen}
\affil{Tuorla Observatory, University of Turku, V\"ais\"al\"antie 20, 
FIN--21500 Piikki\"o, Finland}
\email{jarkot@astro.utu.fi}

\and

\author{A. Treves}
\affil{Universit\`a dell'Insubria, via Valleggio 11, 22100 Como, Italy}
\email{treves@mib.infn.it}

\begin{abstract}

The  correlation between black hole mass $M_{BH}$ and stellar 
velocity dispersion $\sigma$ in nearby elliptical galaxies affords a novel 
way to determine  $M_{BH}$ in active galaxies. 
We report on measurements  of $\sigma$ 
from optical spectra of  7 BL Lac host galaxies. 
The derived values of $\sigma$ are in the range 
of 160 -- 290 km s$^{-1}$ corresponding to $M_{BH}$ of 5 $\times$ 10$^7$ to 
1 $\times$ 10$^9$ M$_\odot$. The average ratio of $M_{BH}$ to the host 
galaxy mass is  1.4 $\times$ 10$^{-3}$, consistent with that  estimated in other 
active and inactive galaxies. 
The velocity dispersions and the derived values of 
$M_{BH}$ of the BL Lacs are similar to those obtained for 
low redshift radio galaxies, in good agreement with the predictions of the 
unified models for radio--loud active galaxies. 

\end{abstract}


\section{Introduction}

The mass of the central black hole (BH) is of paramount importance in 
theoretical models of AGN. In particular, the dependence of BH mass 
($M_{BH}$) on the global host galaxy properties provides clues to the role of 
BHs in galaxy formation and evolution. Dynamical determination of $M_{BH}$ in 
AGN is difficult, because of the bright emission from the nucleus. The main 
method that has proved to be successful for AGN is reverberation mapping of 
broad emission lines, which is extremely time consuming and gives results on 
$M_{BH}$ that depend on the assumed geometry of the accretion disk. 
Therefore, only for a few well studied quasars and Seyfert galaxies $M_{BH}$ 
is known (see e.g. Kaspi et al. 2000; Nelson 2000; Wandel 2002 and references 
therein). Reverberation mapping cannot be employed for BL Lac objects because 
they lack prominent broad emission lines, therefore other methods need to be 
applied. The discovery of a correlation between 
$M_{BH}$ and the luminosity of the bulge in nearby early-type galaxies (e.g. 
Magorrian et al. 1998) offered a new tool for evaluating $M_{BH}$ (see the 
recent reviews by Merritt \& Ferrarese 2001 [hereafter MF01]; 
Kormendy \& Gebhardt 2001). This correlation has been applied so far for a 
sample of nearby quasars (McLure \& Dunlop 2001) and BL Lacs 
(Treves et al. 2001).

Recently, a stricter correlation was found relating $M_{BH}$ with the stellar 
velocity dispersion $\sigma$ of the spheroidal component in nearby inactive 
galaxies (Gebhardt et al. 2000; Ferrarese \& Merritt 2000). This relationship 
clearly demonstrates a connection between BHs and bulges of galaxies and has 
spurred a substantial effort in theoretical modelling (e.g. 
Silk \& Rees 1998; H\"ahnelt \& Kauffmann 2000; 
Adams, Graff \& Richstone 2001). The 
relationship appears to predict more accurately $M_{BH}$, but requires the 
measurement of $\sigma$ in the host galaxies of AGN that is difficult to 
obtain, in particular for objects at moderate or high redshift and with very 
luminous nuclei. On the other hand, BL Lacs have relatively fainter nuclei 
than quasars (e.g. Falomo et al. 1999), and for them this measurement (at 
least for low redshift objects) can be secured with a single spectrum 
observable with a medium-sized telescope.

We present here medium resolution optical spectroscopy of the host
galaxies of 7 BL Lac objects from which we derive the stellar 
velocity dispersion.
According to the shape of their spectral energy distribution (SED), BL Lacs
are broadly distinguished into two types (see Giommi \& Padovani 1995) : 
those  whose SED peaks at 
near-infrared/optical and the $\gamma$-ray MeV regions (low frequency 
peaked BL Lacs or LBL), and those  
that have SED peaking in the UV/X-ray and the $\gamma$-ray TeV energies 
(called high frequency peaked BL Lacs or HBL). 
Our selection of  nearby (z $<$ 0.2) BL Lacs includes 5 HBL and 2 LBL.
For all observed targets  high quality images have
been obtained either from the ground (Falomo \& Kotilainen 1999) or
with HST + WFPC2 (Urry et al. 2000; Falomo et al. 2000). From these
images, the characterization of the host galaxies and the nuclear
luminosity can  be obtained. 

\section{Observations and data analysis}

The observations were obtained in June 2001 using the 2.5m Nordic
Optical Telescope (NOT) equipped with 
ALFOSC\footnote{See URL http://www.not.iac.es for instrument characteristics}. 
Spectra were secured
using two grisms to cover the spectral ranges 4800 -- 5800 \AA~(setup A) 
and 5700 -- 8000 \AA ~(setup B)
at 0.54 \AA~pixel$^{-1}$ and 1.3 \AA ~pixel$^{-1}$ dispersion, respectively. 
This allows us to measure the absorption lines of H$\beta$ (4861 \AA), Mg I
(5175 \AA), Ca E-band (5269 \AA), Na I (5892 \AA) and the TiO + CaI
(6178 \AA), TiO + FeI (6266 \AA) and other absorption line blends from
the host galaxies at a spectral resolution R $\sim$3000.

The chosen grisms combined with a 1\arcsec\ slit yield a spectral resolution 
for velocity dispersion measurement of $\sim$60 -- 80 km s$^{-1}$, which is 
adequate for the expected range of $\sigma$ in luminous ellipticals (e.g. 
Djorgovski \& Davis 1987; Bender, Burstein \& Faber 1992) 
such as the hosts of BL Lacs. 
In addition to the spectra of the BL Lac hosts, we acquired spectra of bright 
stars of type G8-III to K1-III, that exhibit a low rotational velocity 
(V$\times$$\sin{(i)}<$ 20 km$s^{-1}$). These are used as templates of zero 
velocity dispersion. Furthermore, spectra of the well studied nearby 
elliptical galaxy NGC 5831 were also secured in order to provide a test of 
the adopted procedure to derive $\sigma$.

During the observations, seeing ranged between 1\arcsec\ and 1.5\arcsec . \ 
The targets were centered into the slit or positioned 1\arcsec\ away from the 
nucleus and then the 1D spectrum was extracted from an aperture of 
3\arcsec\ - 5\arcsec\ diameter, which is in all cases within the 
effective radius of the host galaxy. 
In one case (Mkn 180), spectra with the object both centered into the slit 
and off-centered by 1\arcsec\ were taken but no significant difference 
was apparent in the shape of the spectral features.  
Standard data reduction was applied to the spectra using the tasks available 
in the {\sc IRAF}\footnote{{\sc IRAF} is distributed by the 
National Optical Astronomy Observatories, which are operated by the 
Association of Universities for Research in Astronomy, Inc., under 
cooperative agreement with the National Science Foundation.} package. 
The procedure includes bias subtraction, flat-fielding, wavelength 
calibration and extraction of 1D spectra. For each observation, we took two 
spectra and combined them in order to remove cosmic ray hits and other 
occasional spurious signals in the detector. In Table 1 we report the list of 
our targets together with the instrumental setup and the S/N of each 
spectrum derived 
from the continuum in the middle of the observed spectral range.

The stellar velocity dispersion $\sigma$ was determined using the 
Fourier Quotient method (e.g. Sargent et al. 1977) implemented in the 
{\sc IRAF} STSDAS package. The spectra were first normalized by subtracting 
the continuum, converted to a logarithmic scale and then multiplied by a 
cosine bell function that apodizes 10\% of the pixels at each end of the 
spectrum. Finally, the Fourier Transform of the galaxy spectra was divided by 
the Fourier Transform of template stars and $\sigma$ was computed from a 
$\chi^{2}$ fit with a Gaussian broadening function (see Bertola et al. 1984; 
Kuijken \& Merrifield 1993 for further details on this method). The $r.m.s.$ 
scatter of the $\sigma$ results using different template stars was typically 
$\sim$10 km$s^{-1}$ and can be considered as the minimum uncertainty of the 
measurement. The observed values of $\sigma$ and their estimated errors are 
reported in the last column of Table 1.

For three objects we have spectra in both spectral ranges. The resulting 
stellar velocity dispersions are in all cases in good agreement, with average 
difference 12 km s$^{-1}$, ensuring sufficient homogeneity of data taken with 
different grisms and/or resolution. Note, however, that there is a tendency 
for the red grism (lower resolution) data to result in slightly larger value 
of $\sigma$. For the nearby elliptical NGC 5831 we obtained 
$\sigma$ = 167$\pm$5 and 185$\pm$10 km s$^{-1}$ for the setup A and B, 
respectively.  
These values are in good agreement with previous measurements in the 
literature ($<\sigma>$ = 168 km s$^{-1}$; Prugniel et al. 1998).  
In Fig. 1 we show an example of the spectrum of the BL Lac object Mrk 501 
compared with that of the elliptical galaxy NGC 5831 observed in both 
spectral ranges.

Since early-type galaxies exhibit some gradients in velocity dispersion 
(Davies et al. 1983; Fisher, Illingworth \& Franx 1995), the measured value 
of $\sigma$  depends somewhat on the distances of the galaxies 
and the size of the used aperture.  In order to compare our values of 
$\sigma$ with the data available in literature (in particular with the MF01 
relationship), we applied aperture corrections according to the procedure 
given in J\"orgensen, Franx \& Kjaergaard (1995). The individual measurements 
of $\sigma$ are therefore corrected to a circular aperture with a metric 
diameter of 1.19$h^{-1}$ kpc, equivalent to 3.4\arcsec\ at the distance of 
the Coma cluster to derive central velocity dispersion $\sigma_c$, 
given in  column 3 of  Table 2.  When 
measurements of $\sigma$ in two spectral ranges are available, 
the average  is reported.

No previous systematic study of the stellar velocity 
dispersion in BL Lacs is available although recently 
Barth, Ho \& Sargent (2002) report optical spectroscopy  
for  Mrk 501. These authors  measured  a value of 
$\sigma$ = 372$\pm$18 km s$^{-1}$ which differs significantly from ours  
($\sigma$ = 291$\pm$13 km s$^{-1}$). 
Their measurement of $\sigma$, derived from 
the region 5200 - 5600 \AA, could be reconciled with 
our value, given their large scatter (81 km s$^{-1}$) fitting the 
data in this range. On the other hand the higher 
$\sigma$ value obtained from the Ca II triplet 
lines (8498, 8542 and 8662 \AA), that are partly blended with telluric 
absorptions, appears inconsistent with our value within the estimated errors. 
Applying aperture correction to the value of Barth et al the 
difference of $sigma$ becomes even larger (by $\sim$ 15 km s$^{-1}$).

\section{Results and discussion}

We have adopted the relationship between $M_{BH}$ and $\sigma_c$ found for 
nearby early-type galaxies that is based on optical spectroscopy (MF01):

$M_{BH}$ = 1.48$\pm$0.24 $\times$ 10$^8$ ($\sigma$/200)$^{4.65\pm0.48}$ [M$_\sun$] \hfill (1) 

We assume that this relationship is also valid for AGN (see e.g. MF01) and in 
particular for BL Lacs. This is consistent with our imaging studies of 
BL Lacs (Falomo \& Kotilainen 1999; Urry et al. 2000; Falomo et al 2000), 
indicating that all our objects are hosted by luminous ellipticals. 
The derived values of $M_{BH}$ are reported in column 4 of Table 2, 
where the errors are 
the composition in quadrature of uncertainties in $\sigma$ and in the MF01 
relationship. Using the Gebhardt et al. (2000) relationship instead of the 
one by MF01, tends to yield slightly lower values of $M_{BH}$ but does not 
substantially modify our main conclusions. The values of  $M_{BH}$ in Table 2 
span a factor 
$\sim$20 from 5 x 10$^7$ M$_\odot$ for PKS 2201+04 to 9 x 10$^8$ M$_\odot$ for Mrk 501.

As mentioned above, $M_{BH}$ is also correlated, but with a larger scatter, 
with the luminosity of the bulge of the host galaxy. The host galaxy 
absolute magnitude M$_R$ (uncorrected for extinction) and the effective radii R$_e$ of the 
7 BL Lacs are given in columns 5 and 6 of Table 2. 
$M_{BH}$ was thus calculated following the 
relationship by McLure \& Dunlop (2002):

log $M_{BH}$ = --0.50$\pm$0.05 M$_R$ -- 2.91$\pm$1.23 [M$_{\odot}$] \hfill (2)

The corresponding values of $M_{BH}$ are given in column 7 of Table 2. 
For most sources the difference of $M_{BH}$ derived with the two methods is 
within the estimated uncertainty. The average values of $M_{BH}$ for our 
BL Lacs derived, respectively, from $\sigma$ and the host luminosity are: 
$<$log$ M_{BH}>_{\sigma}$ = 8.62 $\pm$ 0.23 and 
$<$log$ M_{BH}>_{host}$ = 8.66 $\pm$ 0.25.

In two cases (I Zw 187 and PKS 2201+04) a factor of $\sim$3 difference in 
$M_{BH}$ is found. We note that for PKS 2201+04 we derive a significantly 
lower velocity dispersion with respect to the rest of the observed sources 
leading to a low $M_{BH}$. On the other hand, for this target $\sigma$ is 
well determined (good S/N data and the two spectral ranges giving similar 
result).

 In our sample there are two  LBL type and five HBL
type BL Lacs (column 2 of Table 2).  With the caveat that the number of
studied objects is very small, we find no significant difference of
$M_{BH}$ between the two types of BL Lacs.
 
The measurements of $\sigma$ combined with the effective radii of the
host galaxies can be used to estimate the mass of the hosts through
the relationship (Bender et al. 1992)  :

$M_{host}$=5$\sigma^{2}r_e$/G   \hfill (3) 

This dynamical mass (column 8 of Table 2) turns out to be in the range
of 1 -- 4 x 10$^{11}$ M$_\odot$. The ratio between $M_{BH}$ and
$M_{host}$ is in the range of 0.5 -- 3.6 $\times$ 10$^{-3}$, with
$<M_{BH}$/$M_{host}>$ = 1.4 $\times$ 10$^{-3}$. This is in good
agreement with values derived for both AGN and inactive galaxies
($<M_{BH}$/$M_{host}>$ = 1.2 $\times$ 10$^{-3}$ ; McLure \& Dunlop
2001; MF01).

In the unified model of radio-loud AGN, BL Lacs are believed to 
be drawn from the 
population of radio galaxies according to orientation effects (e.g. 
Urry 1999). It is therefore interesting to compare orientation-independent 
properties, such as the velocity dispersion of the host galaxy, of 
BL Lac and radio galaxy populations. 
We show in Fig. 2 the comparison of the host luminosity 
M$_R$ vs. log$\sigma$ (the Faber-Jackson relationship) for the BL Lacs with 
respect to a large sample of low redshift radio galaxies 
(Bettoni et al. 2001). Both samples follow quite well the original 
Faber \& Jackson (1976) correlation. The similarity of the distributions of 
$\sigma$ for these two samples (Fig. 3) furthermore implies that the 
distributions of $M_{BH}$ in radio galaxies and BL Lacs are 
indistinguishable, consistent with the model that both types of AGN belong to 
the same population but are observed from different orientation angles. 

\begin{acknowledgments}

We thank D. Bettoni for helpful discussions and suggestions on the 
measurements of the velocity dispersion.  This work has received partial 
support under contracts COFIN 2001/028773 and ASI-IR-35. 
Nordic Optical Telescope is operated on the island of La Palma jointly 
by Denmark, Finland, Iceland, Norway, and Sweden, in the Spanish 
Observatorio del Roque de los Muchachos of the Instituto de Astrofisica 
de Canarias. We thank the NOT staff for kind hospitality and support. 
The data presented here have been taken using ALFOSC, which is owned by 
the Instituto de Astrofisica de Andalucia (IAA) and operated at the 
Nordic Optical Telescope under agreement between IAA and the NBIfAFG of 
the Astronomical Observatory of Copenhagen.

\end{acknowledgments}


\begin{figure*}
\epsscale{1}
\plotone{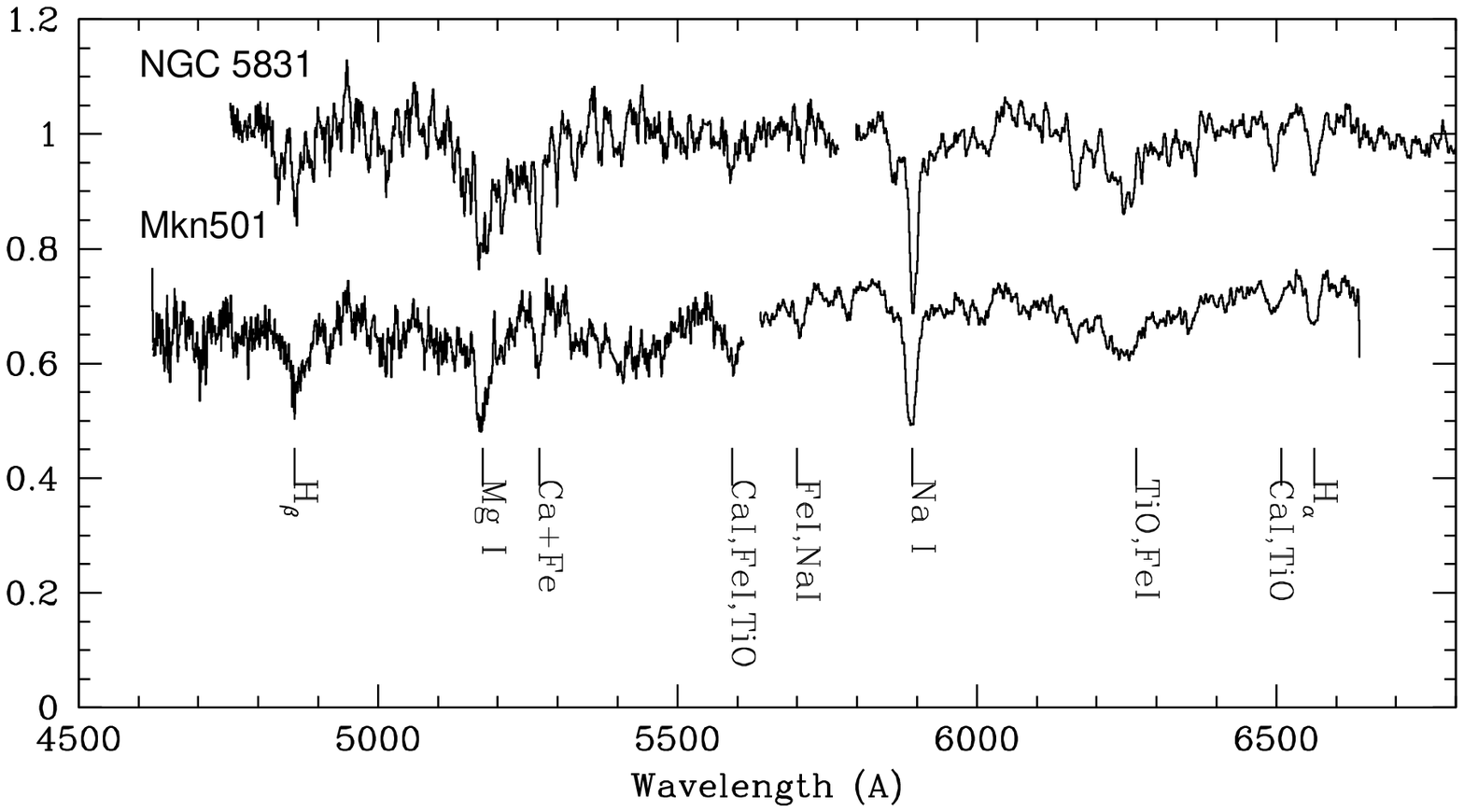}
\caption{Optical spectra of the BL Lac object Mrk 501 (z = 0.034) and of the 
nearby elliptical galaxy NGC 5831 (z = 0.0055). The spectra are normalized to 
the continuum and plotted in the rest frame.}
\end{figure*}

\clearpage

\begin{figure}
\epsscale{1}
\plotone{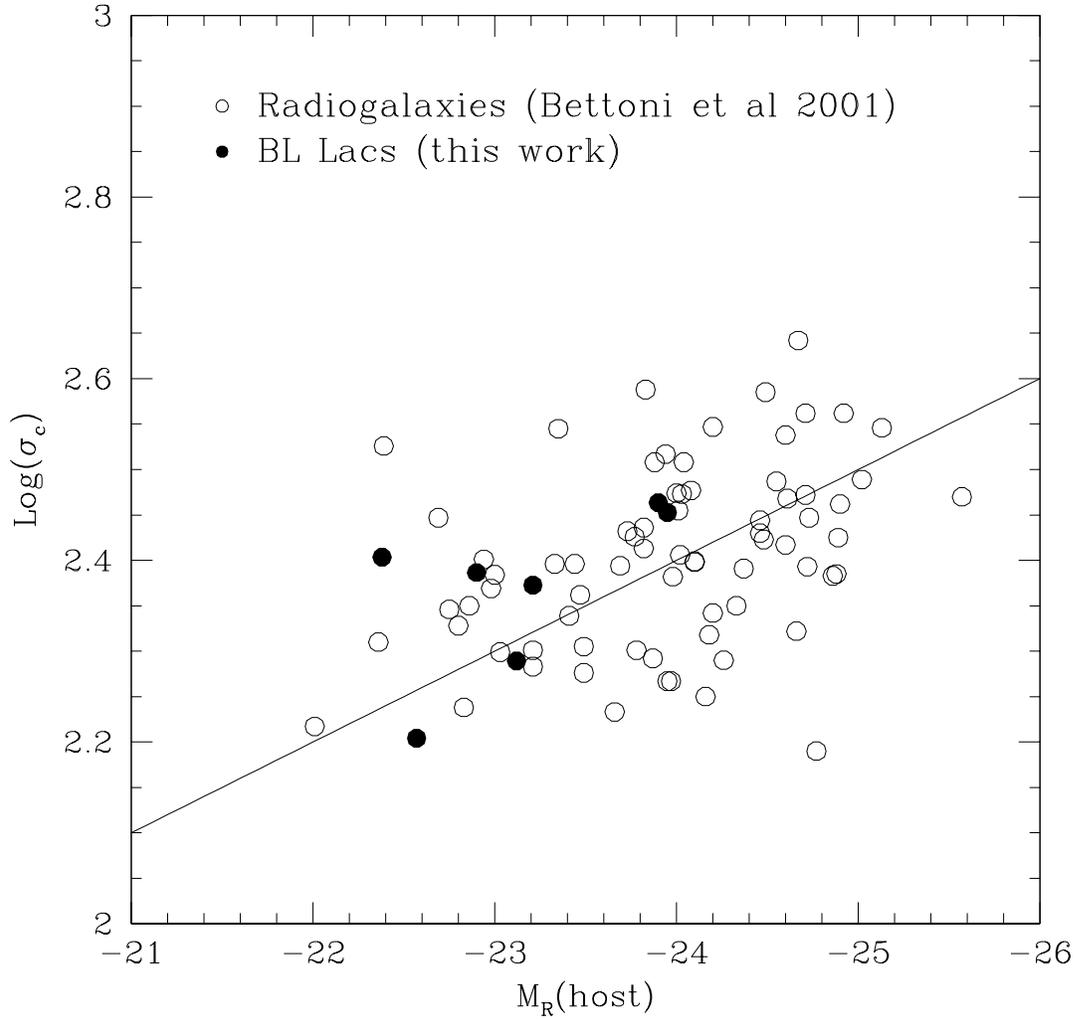}
\caption{The host galaxy stellar velocity dispersion $\sigma_c$ vs. the 
$R$-band absolute magnitude for the BL Lacs (filled circles) and 
for low redshift radio galaxies (Bettoni et al. 2001; 
open circles). The solid line indicates the original Faber \& Jackson (1976) 
relationship, transformed into the $R$-band.}
\end{figure}

\clearpage

\begin{figure}
\epsscale{1}
\plotone{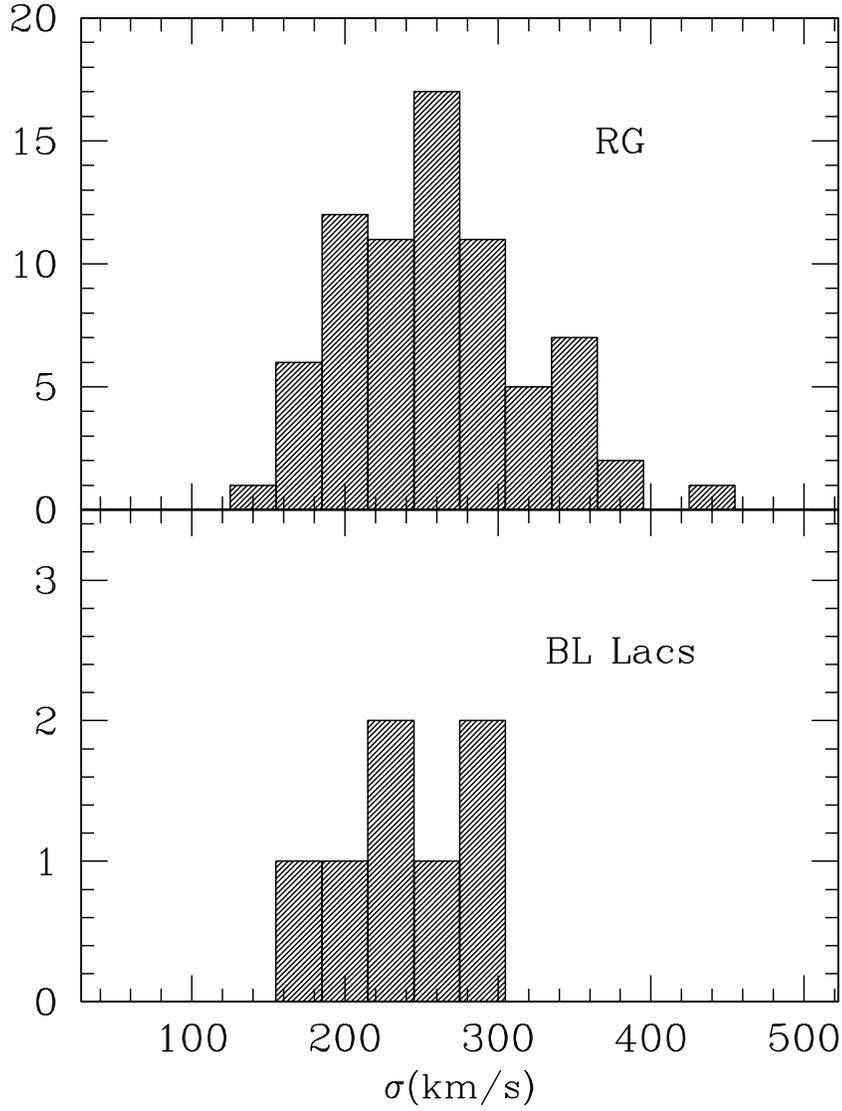}
\caption{The distribution of the stellar velocity dispersion $\sigma_c$ of 
low redshift radio galaxies (Bettoni et al. 2001; {\it upper panel}) compared 
with that of the BL Lacs studied in this paper ({\it lower panel}).}
\end{figure}

\clearpage

%
\begin{deluxetable}{llrlrll}
\tablecolumns{7} 
\tablewidth{0pc} 
\tablecaption{Journal of observations and results.}
\tablehead{
\colhead{Object} & \colhead{z} & \colhead{Setup}& 
\colhead{Exposure} &\colhead{S/N} & \colhead{$\sigma$}\\
\colhead{} & \colhead{} & \colhead{}  & \colhead{(sec)} & \colhead{} & 
\colhead{km s$^{-1}$}}
\startdata
NGC 5831	& 0.0055 & A & 600  &45  & 167$\pm$5 \\
	        &        & B & 600  &55  & 185$\pm$10 \\
                &        &    &      &    &            \\
Mrk 421		& 0.031	 &  B & 3600 &40  & 220$\pm$10 \\
Mrk 180		& 0.045	 &  A & 3600 &40  & 225$\pm$10 \\
Mrk 501		& 0.034	 &  A & 2400 &40  & 265$\pm$10 \\
       		&      	 &  B & 3600 &55  & 280$\pm$15 \\
I Zw 187 	& 0.055  &  B & 3600 &30  & 253$\pm$15 \\
3C 371 		& 0.051  &  A & 2400 &40  & 255$\pm$15 \\
       		&        &  B & 3600 &35  & 265$\pm$20 \\
1ES 1959+65 	& 0.048  &  B & 2400 &18  & 180$\pm$15 \\
PKS 2201+04 	& 0.027  &  A & 3600 &30  & 148$\pm$5 \\
            	&        &  B & 2400 &50  & 153$\pm$8 \\
   \enddata
  \end{deluxetable}

\begin{deluxetable}{llllllll}
\tablecolumns{8} 
\tablewidth{0pc} 
\tablecaption{Velocity dispersion and BH masses of BL Lacs}
\tablehead{
\colhead{Object} & \colhead{LBL/HBL} & \colhead{$\sigma_c$} & 
\colhead{log($M_{BH}$)$_\sigma$} & \colhead{M$_R$} & \colhead{R$_e$} & 
\colhead{log($M_{BH}$)$_{bulge}$} & \colhead{log($M_{(host)}$)} \\
\colhead{} & \colhead{} & \colhead{km s$^{-1}$} & 
\colhead{[M$_{\odot}$]} & \colhead{} & \colhead{kpc} & 
\colhead{[M$_{\odot}$]} & \colhead{[M$_{\odot}$]} 
}
\startdata
Mrk 421     & HBL &  236$\pm$10& 8.50$\pm$0.18 & -23.12 &3.4& 8.65 & 11.20 \\
Mrk 180     & HBL &  244$\pm$10& 8.57$\pm$0.19 & -22.81 &5.0& 8.50 & 11.45 \\
Mrk 501     & HBL &  291$\pm$13& 8.93$\pm$0.21 & -23.87 &15 & 9.00 & 11.59 \\
I Zw 187    & HBL &  253$\pm$15& 8.65$\pm$0.18 & -22.22 &4.7& 8.20 & 11.39 \\
3C 371      & LBL &  284$\pm$18& 8.88$\pm$0.20 & -23.67 &2.9& 8.90 & 11.32 \\
1ES 1959+65 & HBL &  195$\pm$15& 8.12$\pm$0.13 & -22.48 &6.6& 8.30 & 11.27 \\
PKS 2201+04 & LBL &  160$\pm$7 & 7.72$\pm$0.13 & -22.36 &5.1& 8.27 & 11.00 \\
   \enddata
  \end{deluxetable}

\end{document}